\documentclass[journal=nalefd,manuscript=letter]{achemso}

\usepackage[T1]{fontenc}       
\usepackage{graphicx}

\author{Peter Schnauber}
\author{Johannes Schall}
\author{Samir Bounouar}
\affiliation[TU Berlin]
{Institut f\"ur Festk\"orperphysik, Technische Universit\"at Berlin, Hardenbergstra{\ss}e 36, 10623 Berlin, Germany}
\author{Theresa H\"ohne}
\affiliation[ZIB]{Zuse Institute Berlin, Takustra{\ss}e 7, 14195 Berlin, Germany}
\author{Suk-In Park}
\author{Geun-Hwan Ryu}
\affiliation[Korea]{Center for Opto-Electronic Material and Devices Research, Korea Institute for Science and Technology, Hwarangno 14-gil 5, Seongbuk-gu, Seoul 02-791, Republic of Korea}
\author{Tobias Heindel}
\affiliation[TU Berlin]
{Institut f\"ur Festk\"orperphysik, Technische Universit\"at Berlin, Hardenbergstra{\ss}e 36, 10623 Berlin, Germany}
\author{Sven Burger}
\affiliation[ZIB]{Zuse Institute Berlin, Takustra{\ss}e 7, 14195 Berlin, Germany}
\author{Jin-Dong Song}
\affiliation[Korea]{Center for Opto-Electronic Material and Devices Research, Korea Institute for Science and Technology, Hwarangno 14-gil 5, Seongbuk-gu, Seoul 02-791, Republic of Korea}
\author{Sven Rodt}
\author{Stephan Reitzenstein}
\affiliation[TU Berlin]
{Institut f\"ur Festk\"orperphysik, Technische Universit\"at Berlin, Hardenbergstra{\ss}e 36, 10623 Berlin, Germany}
\email{stephan.reitzenstein@physik.tu-berlin.de}

\title[An \textsf{achemso} demo]
  {Deterministic integration of quantum dots into on-chip multi-mode interference beamsplitters using in-situ electron beam lithography}

\abbreviations{IR,NMR,UV}

\begin{document}


\begin{abstract}
The development of multi-node quantum optical circuits has attracted great attention in recent years. In particular, interfacing quantum-light sources, gates and detectors on a single chip is highly desirable for the realization of large networks. In this context, fabrication techniques that enable the deterministic integration of pre-selected quantum-light emitters into nanophotonic elements play a key role when moving forward to circuits containing multiple emitters. Here, we present the deterministic integration of an InAs quantum dot into a 50/50 multi-mode interference beamsplitter via in-situ electron beam lithography. We demonstrate the combined emitter-gate interface functionality by measuring triggered single-photon emission on-chip with $g^{(2)}(0) = 0.13\pm 0.02$. Due to its high patterning resolution as well as spectral and spatial control, in-situ electron beam lithography allows for integration of pre-selected quantum emitters into complex photonic systems. Being a scalable single-step approach, it paves the way towards multi-node, fully integrated quantum photonic chips.
\end{abstract}

\section{Main}
On-chip quantum optical circuits offer superior performance and scalability compared to bulky optical setups\cite{Politi2008,Harris2017}. Additionally, the up-scaling of quantum systems is quintessential for the realization of photonic quantum computers that exceed the calculation capabilities of their classical counterparts\cite{Ladd2010}. Aiming at circuits that involve many quantum emitters, gates and detectors, it is desirable to integrate all these elements onto a single chip to enhance the functionality and avoid losses between different technology platforms\cite{Aspuru-Guzik2012,Silverstone2016,Reithmaier2015}. Against this background, the deterministic integration of single quantum emitters into on-chip quantum nanophotonic elements has gained a lot of attention recently. Deterministic integration techniques range from the positioned growth of quantum emitters\cite{Calic2017,Strauss2017,Schneider2012,Joens2013}, marker based alignment of nanostructures to emitters\cite{Sapienza2015,Coles2016}, over low temperature in-situ lithography\cite{Dousse2008,Sartison2017,Gschrey2013} to AFM transfer methods of semiconductor quantum devices onto a separate chip\cite{Zadeh2016,Kim2017}.

Here, we report on the deterministic integration of a pre-selected single InAs quantum dot into an on-chip 50/50 multi-mode interference (MMI) beamsplitter via in-situ electron beam lithography (EBL). The combined emitter-waveguide interface functionality is proven through an on-chip Hanbury-Brown and Twiss (HBT) experiment confirming the on-chip guiding and beamsplitting of single-photons. In this context, our in-situ electron beam lithography approach is ideally suited since it has high lateral resolution and features design flexibility enabling the patterning of complex structures aligned to pre-selected QDs. Moreover, being a scalable, spectrally and spatially selective single-step EBL method\cite{Gschrey2015b,Gschrey2015} it paves the way for integrating multiple semiconductor quantum emitters at a specified wavelength into large on-chip photonic networks that rely on precise spectral and spatial control of the emitters. At the same time, it is not subject to complex alignment procedures as in marker based approaches\cite{Sapienza2015,Coles2016}. Also, the in-situ EBL is conducted in a single step avoiding for instance complicated AFM tip transfer methods\cite{Zadeh2016,Kim2017}.

In general, on-chip quantum optical circuits comprise emitters, waveguide (WG) sections, gates and detectors\cite{OBrien2009,Lodahl2015}. To show the high potential of our deterministic nanotechnology platform we deterministically integrate a QD into an on-chip 50/50 beamsplitter $-$ the key building block for quantum gates. Hereby, we interface the emitter and beamsplitter section on a single chip. As 50/50-splitter we choose a robust multi-mode interference coupler\cite{Soldano1995} in contrast to the widely used evanescent couplers\cite{Politi2008,Prtljaga2014,Joens2015,Rengstl2015}. The coupling ratio in evanescent couplers is strongly dependend on the emitter wavelength and the gap size between the coupler arms. Here, it becomes difficult to control the gap size and as a result the splitting ratio, since the etch rate during plasma etching is reduced in small gaps due to a decreased material transport. Furthermore, evanescent couplers include four bend sections (one arm) that introduce additional optical loss. In contrast to this, MMI couplers rely on the self-imaging effect. That is, when light is launched from a single-mode access waveguide into a sufficently large multi-mode section, \textit{N}-fold self-images of the access waveguide are formed at characteristic distances $D_{N}$ inside the multi-mode section\cite{Ulrich1975}. Placing single-mode exit waveguides at the locations of these self-images allows one to realize efficient beamsplitting devices. Noteworthy, MMI couplers with tapered access and exit waveguides benefit from relaxed fabrication tolerances, with the easy-to-control width of the multi-mode section being the most crucial lithography parameter\cite{Soldano1995}. In addition, footprints of evanescent couplers and MMI couplers are of comparable sizes. Considering quantum optic circuits, MMI couplers have already been proven to enable quantum interference effects involving up to four photons\cite{Peruzzo2011}. Therefore, MMI couplers represent an attractive option for future realizations of quantum optical gates.

\begin{figure}
\centering
\includegraphics[width=8 cm]{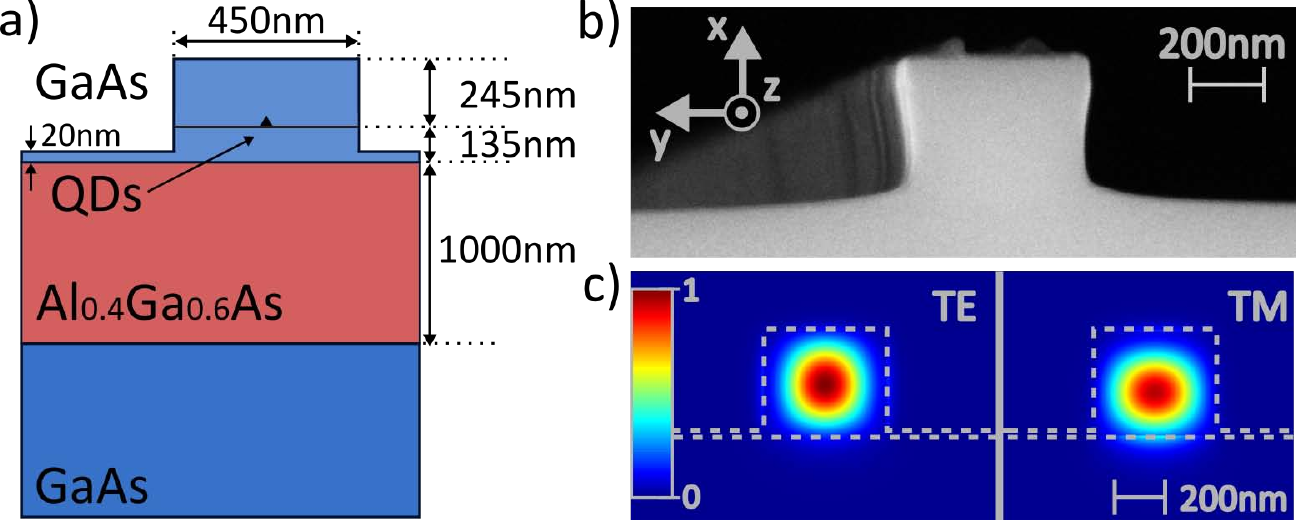}
  \caption{a) Schematic view of the sample geometry b) SEM image of a cleaved waveguide facet. c) Visualization of simulated electric field intensity ($|\vec{E}|^2$) distribution of fundamental WG modes of a 450\,nm wide waveguide.}
  \label{fgr:sample}
\end{figure}

\begin{figure}
\centering
\includegraphics[width=16 cm]{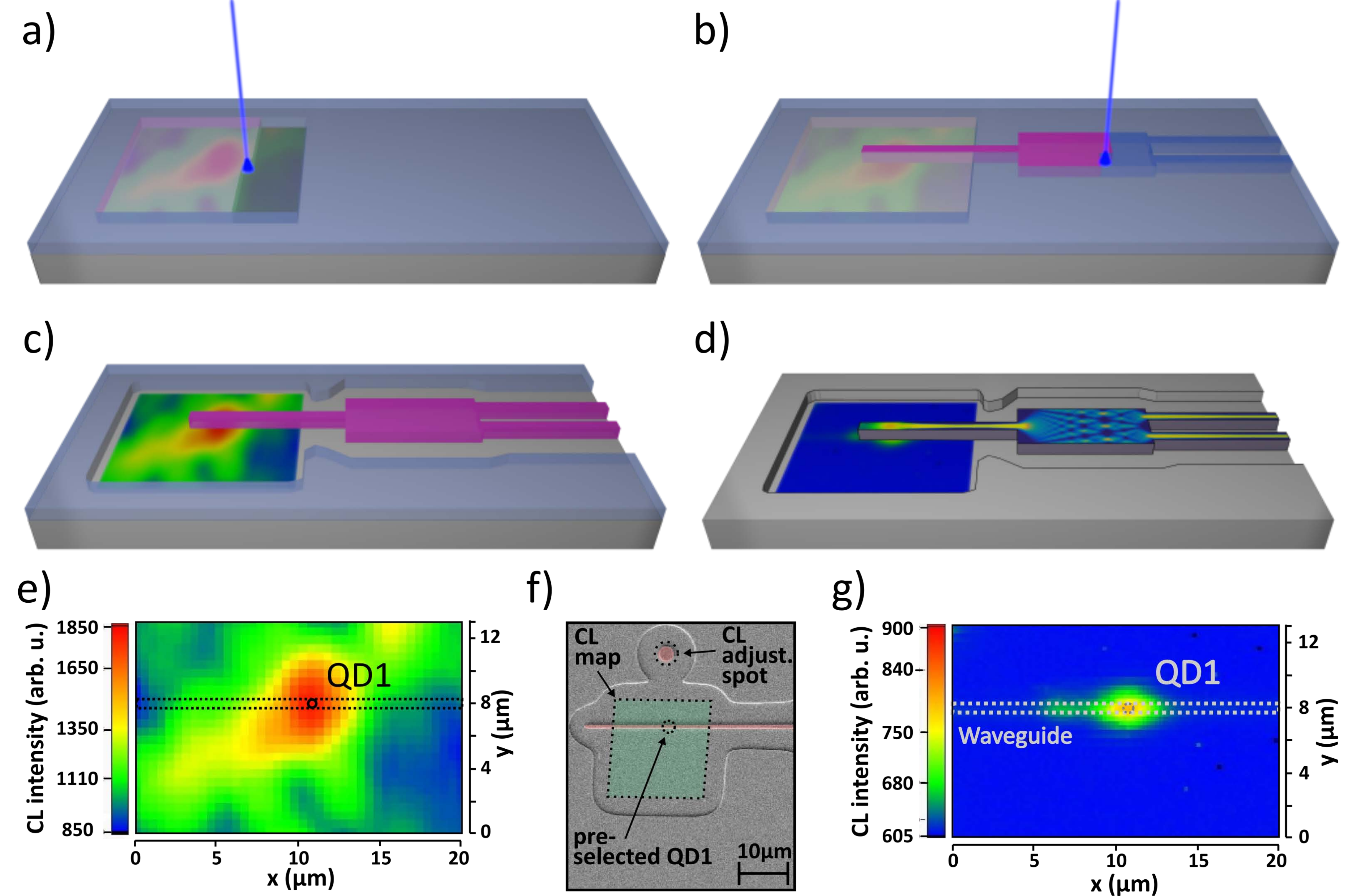}
  \caption{a) - d) Schematic visualization of the in-situ EBL device manufacturing process: a) Suitable QDs are located by taking CL maps at 7\,K (exposed resist becomes soluble). b) A waveguide accessing a photonic element is patterned at the location of a target QD at 7\,K (patterned resist becomes unsoluble). c) Soluble resist is removed during developing. Due to the proximity effect, a margin of resist around the microstructure and CL map is also soluble. d) The resist mask is transferred into the semiconductor by anisotropic etching and residual resist is removed. A CL map taken after sample processing and an exemplary MMI field distribution visualize the successful integration of the target QD into an on-chip photonic element.  e) CL map before EBL patterning. Well-localized emission from the QD is observed (spectral filter from 894.3\,nm to 903.7\,nm, pixel size 500\,nm\,$\times$\,500\,nm). The dotted lines indicate the area of the subsequent waveguide patterning. f) False-color SEM image of the fully processed WG containing QD1. The spot over the map was used to adjust the light-collection optics for maximum CL signal. The green area indicates where the CL map was taken. g) CL map of QD1 after patterning and processing the sample. The emission of QD1 from inside the WG to the top is well resolved (spectral filter from 896\,nm to 909\,nm, pixel size 250\,nm\,$\times$\,250\,nm). A weak signal from a second QD can be seen on the left side.}
  \label{fgr:Maps}
\end{figure}

For the fabrication of our samples we adapt the layer design reported in Refs.\,[25,\,28] (see Fig.\,\ref{fgr:sample}\,a). The sample consists of a GaAs waveguide core and an AlGaAs cladding layer, allowing for reliable growth and straightforward processing of the sample. A low density ($< 10^{7}\, \frac{1}{\textrm{cm}^2}$) InAs QD-layer is grown at the maximum of the expected fundamental TE waveguide mode (see Fig.\,\ref{fgr:sample}\,c). Manufacturing $w=450\,$nm wide single-mode waveguides (Fig.\,\ref{fgr:sample}\,b) by etching $h=360\,$nm into the GaAs, we calculate that a fraction of about $\beta=14\,\%$ of the QD emission is coupled into the fundamental waveguide mode at a wavelength of 900\,nm, 7\,\% for each propagation direction, see \textit{Methods}. Nanophotonic elements are defined at the location of pre-selected quantum dots using in-situ EBL: The samples are covered with a dual-tone resist\cite{Kaganskiy2016} and mounted onto a liquid helium flow cryostat inside a scanning electron microscope (SEM), which contains a cathodoluminescence (CL) spectroscopy extension. In a first step, CL maps are recorded by scanning the electron beam over the sample at a temperature of 7\,K, see Fig.\,\ref{fgr:Maps}\,a and \ref{fgr:Maps}\,e. Waveguides and MMI splitters are structured at the location of target single-QDs immediately afterwards, see Fig.\,\ref{fgr:Maps}\,b, creating an unsoluble resist mask. Due to the proximity effect, resist within 5\,$\mu$m of the patterned area becomes soluble (positive-tone regime). After patterning, the sample is unmounted and soluble resist is removed during development (see Fig.\,\ref{fgr:Maps}\,c). Finally, the pattern is transferred into the GaAs by dry etching, resulting in a monolithic device (see Fig.\,\ref{fgr:Maps}\,d and \ref{fgr:Maps}\,f). Since the CL mapping process already introduces an electron dose of 8\,$\frac{\textrm{mC}}{\textrm{cm}^2}$, special care is required to match the width of the waveguide sections lying inside the mapping region with the ones outside of it, avoiding unwanted light scattering. Moreover, if a waveguide is terminated in close proximity to the QD, interference effects between the mode propagating in $+z$-direction with the reflected $-z$-mode affect the coupling of the QD emission to the waveguide\cite{Hoehne2017}. In order to avoid these effects, we terminate the waveguide more than $10\,\mu$m away from the emitter. The liquid helium flow cryostat causes a small sample drift of at least 14\,nm per minute\cite{Gschrey2015}. All EBL patterns and their respective scanning routines are optimized such that the influence of the drift is minimized and smooth transitions between all photonic structures are guaranteed. More information on the in-situ EBL technology platform is found in \textit{Methods}. After the device is processed the successful integration of target QDs is evaluated and confirmed through CL maps, see Fig.\,\ref{fgr:Maps}\,g. Clearly, the target QD, from now on dubbed QD1, is positioned inside the waveguide. A weak second emitter can be seen to its left. The spectral and spatial accuracy of our technique has been reported previously\cite{Gschrey2015}. 

We further characterize the deterministic QD-waveguide interface through microphotoluminescence ($\mu$PL) spectroscopy. The samples are cleaved perpendicular to the waveguide axis to allow for optical access to the waveguides from the chip facet, see Fig.\,\ref{fgr:sample}\,b. The chips are mounted onto a liquid helium flow cryostat and cooled down to 10\,K. The pre-selected QD is excited from the top by coupling a $\lambda =785\,$nm off-resonant laser through a microscope objective (MO) onto the sample. A second $\textrm{NA}=0.4$ MO is aligned to the waveguide facet in a 90$^\circ$ geometry to collect the QD emission, which is coupled into the waveguide mode. The QD light is sent into a monochromator and detected with either a charge-coupled device (CCD) camera or single-photon counting modules (SPCMs).

\begin{figure}
\centering
\includegraphics[width=8 cm]{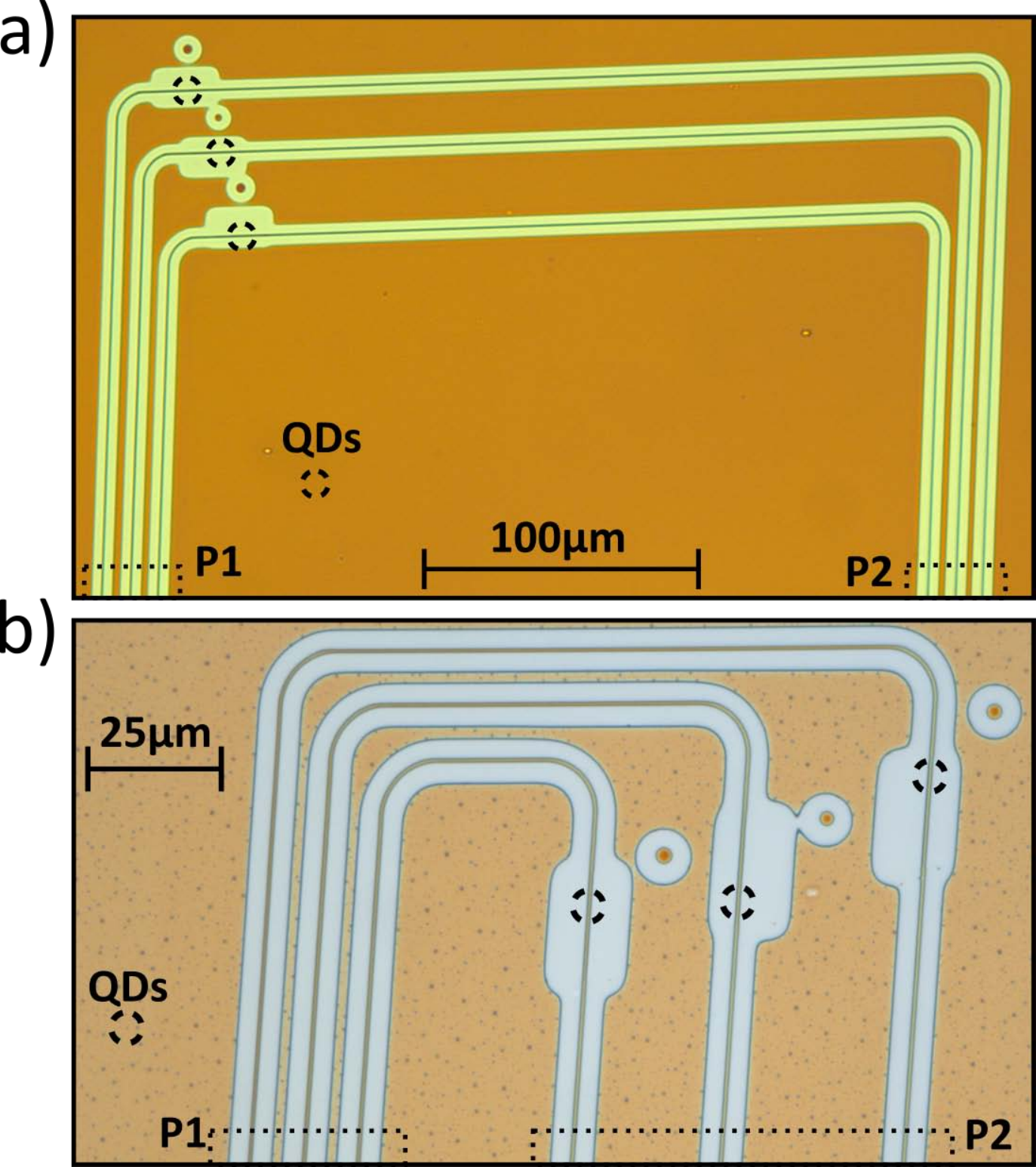}
  \caption{Microscope images of U-shaped waveguide structures for loss characterization. Deterministically integrated QDs, marked by black circles, are located inside the WGs in the mapped areas. a) Propagation loss: The path length difference for photons exiting from P1 and P2 is given by the upper waveguide section. b) Bend loss: Photons exiting from P1 have passed through two bends with radius 10\,$\mu$m, the ones at P2 have passed through none.}
  \label{fgr:Loss}
\end{figure}

Before fabricating MMI couplers, we first characterize the single-mode access and exit waveguides by determining their losses. Having full control of the position of target QDs in on-chip waveguide networks, we use a novel technique to measure the loss that arises from light propagation along straight waveguides and from passing through bent waveguide sections. Previous works have relied on measuring the emission of QDs located at different distances from the facet\cite{Joens2015,Rengstl2015,Reithmaier2013}. Due to random positioning of the QDs inside the waveguide the $\beta$-factor for each quantum dot is not under control, requiring a very large number of inspected QDs within a statistical analysis. In this work, we deterministically integrate target QDs into U-shaped waveguides, see Fig.\,\ref{fgr:Loss}\,a, whose two ends both terminate at the sample facet. Due to the mirror symmetry along and perpendicular to the waveguide axis, the QD emission is coupled equally strong into the fundamental mode travelling in $+z$- and $-z$-direction, independent of the QD position. By placing the QD closer to one end of the U-shaped waveguide, the light propagates a shorter path in side 1, and a longer path in side\,\nopagebreak2 of the U-shaped structure. By comparing the emission intensity at facet 1 to facet 2, we can directly deduce the propagation loss from the path length difference independent of the $\beta$-factor. The loss for passing through a single waveguide bend is determined by placing target QDs close to facet 2 of the U, see Fig.\,\ref{fgr:Loss}\,b. Hence, emission measured at facet 2 has not passed through bends, while emission at facet 1 has travelled through two bend sections, allowing for a straightforward evaluation of bend losses, in which propagation loss is accounted for. Since the two sides of the U-shaped structures show slight variations in terms of waveguide-to-bend interfaces and cleaved facets, we manufacture a series of U-shaped waveguides and average the results. We use waveguides with a width of $450\,$nm, a height of $360\,$nm and a bend radius of 10\,$\mu$m. For the propagation loss $L_{\textrm{Prop}}$, we analyze $N_{\textrm{Prop}}=16$ structures and find $L_{\textrm{Prop}}=(12.1\pm 6.2)\,\frac{\textrm{dB}}{\textrm{mm}}$. Examining $N_{\textrm{Bend}} =10$ structures, we find that when passing through a single bend, $L_{\textrm{Bend}}= (1.11\pm 0.69)\, \textrm{dB}$ of the light is lost. While the bend loss compares well to other works in the same material system\cite{Joens2015,Rengstl2015,Reithmaier2013}, the propagation loss is considerably higher. We attribute the propagation loss mainly to a pronounced sidewall roughness arising from our liquid helium temperature EBL manufacturing process with self-made patterning tools. We expect improved results with professional in-situ EBL patterning machines in the future.

\begin{figure}
\centering
\includegraphics[width=16 cm]{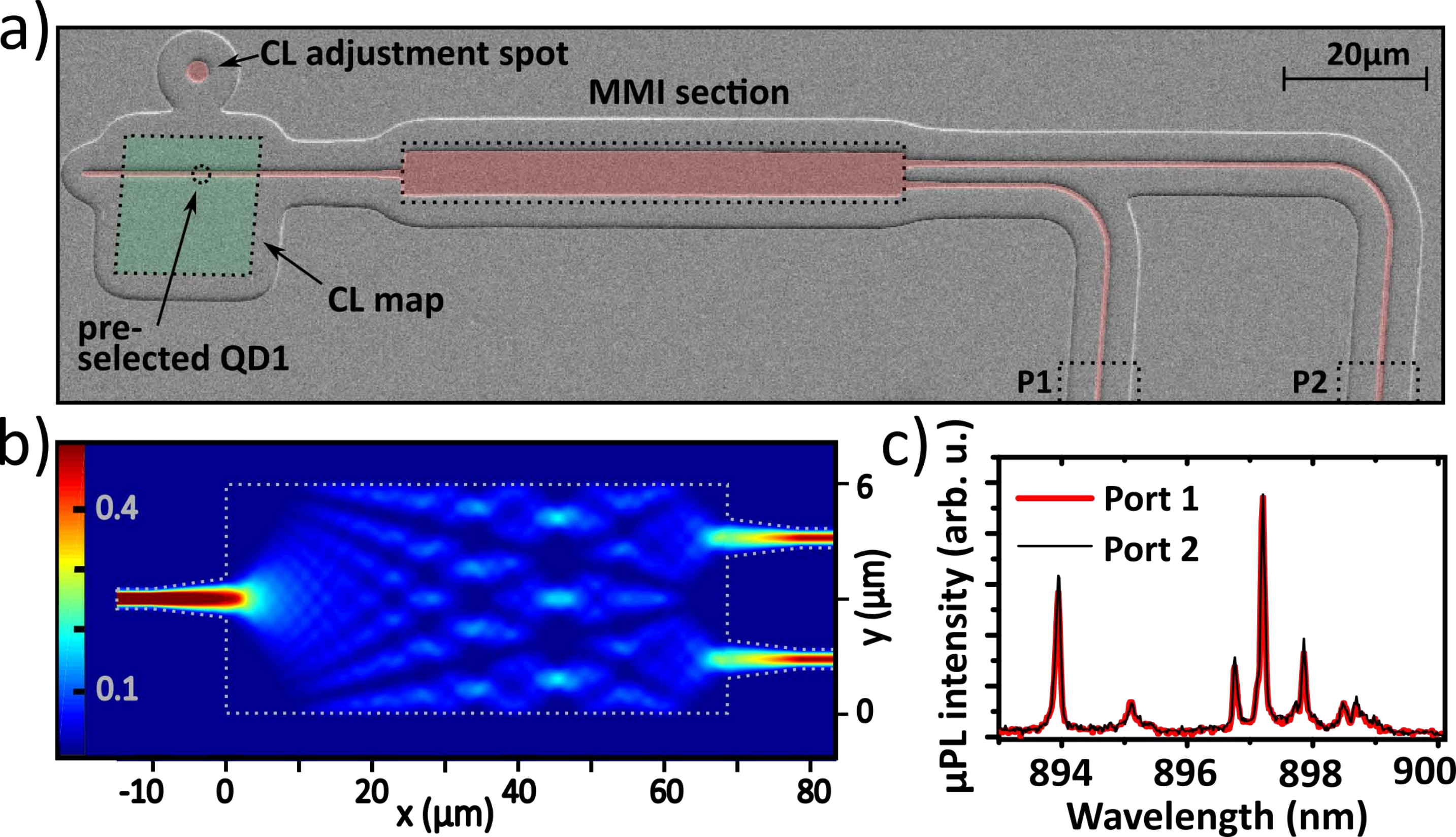}
  \caption{a) False-color SEM image of QD1 integrated into a $1\times 2$ MMI coupler via in-situ EBL. b) Visualization of the simulated electric field intensity distribution ($|\vec{E}|^2$) 245\,nm below the surface in the optimized MMI and taper device (dimensions not to scale). c) $\mu$PL spectra of QD1 taken from port 1 and port 2 of the device shown in a). 50/50 splitting over the whole spectrum is observed.}
  \label{fgr:MMI}
\end{figure}

In order to find a design for the fabrication of MMI couplers which allows for high transmission $T$ we perform numerical simulations of light propagation through the device. Rigorous simulations of light propagation are performed using a finite element method (FEM), see \textit{Methods}. For given waveguide cross-section we scan the width $W_{\textrm{Sim}}$ and length $L_{\textrm{Sim}}$ of the multi-mode section. Maximum transmission of $T_{\textrm{Sim}}=97\,\%$ is obtained for $W_{\textrm{Sim}}=6.05\,\mu$m and $L_{\textrm{Sim}}=68.88\,\mu$m, which compares favourably to the bend losses in evanescent couplers. Fig.\,\ref{fgr:MMI}\,b visualizes the computed field intensity distribution for the optimized design parameters.

Using our in-situ EBL, we integrate QD1 into a tapered access waveguide of a $1\times2$ symmetrical MMI coupler, see Fig.\,\ref{fgr:MMI}\,a. With the help of cross-sectional SEM images, the width $W_{\textrm{Exp}}$ and length $L_{\textrm{Exp}}$ of the device are determined to be $W_{\textrm{Exp}}=(6.05\pm0.02)\,\mu$m and $L_{\textrm{Exp}}=(69.3\pm0.3)\,\mu$m. For more information on the MMI geometry, see the \textit{supporting info S1}. Tapered exit waveguides are defined at the positions of the two-fold self-images. The exit waveguides are equipped with single bend sections in order to separate them by a distance of 40\,$\mu$m. This enhances the spatial separation of the emission from the output arms, while still allowing us to access the emission of both exit waveguides at the sample facet simultaneously with a single 50x MO in a later experiment. Applying $\mu$PL spectroscopy, we excite QD1 with an off-resonant continuous wave (CW) laser and collect the emission of exit port 1 and exit port 2, resulting in the spectra in Fig.\,\ref{fgr:MMI}\,c. The collecting MO is aligned to maximize the detection of the $897.3\,$nm emission line and clear 50/50 splitting is observed over a spectral range of 23\,nm. In addition, we measure the transmission $T_{\textrm{Exp}}$ through an MMI coupler for a series of $N_{\textrm{MMI}}=5$ devices, employing the U-shaped waveguide geometry described above, see \textit{supporting info S2}. The emission intensity from a control port is compared to the sum of the MMI output ports' intensities, corrected by propagation loss outside the MMI and a transmission $T_{\textrm{Exp}}=(61\pm15)\,\%$ is determined. Since the MMI transmission is very robust to imperfections in the length $L_{\textrm{MMI}}$ (see \textit{supporting info S3}), we attribute the difference of the measured transmission $T_{\textrm{Exp}}$ to the simulated $T_{\textrm{Sim}}$ mostly to geometrical deviations of the access and exit waveguides from the optimum geometry as well as sidewall roughness in the 69.3\,$\mu$m long MMI section. Yet, a transmission of $T_{Exp}=61\,\%$ is considerably higher than what can be expected from evanescent couplers consisting of four bend sections.

\begin{figure}
\centering
\includegraphics[width=8 cm]{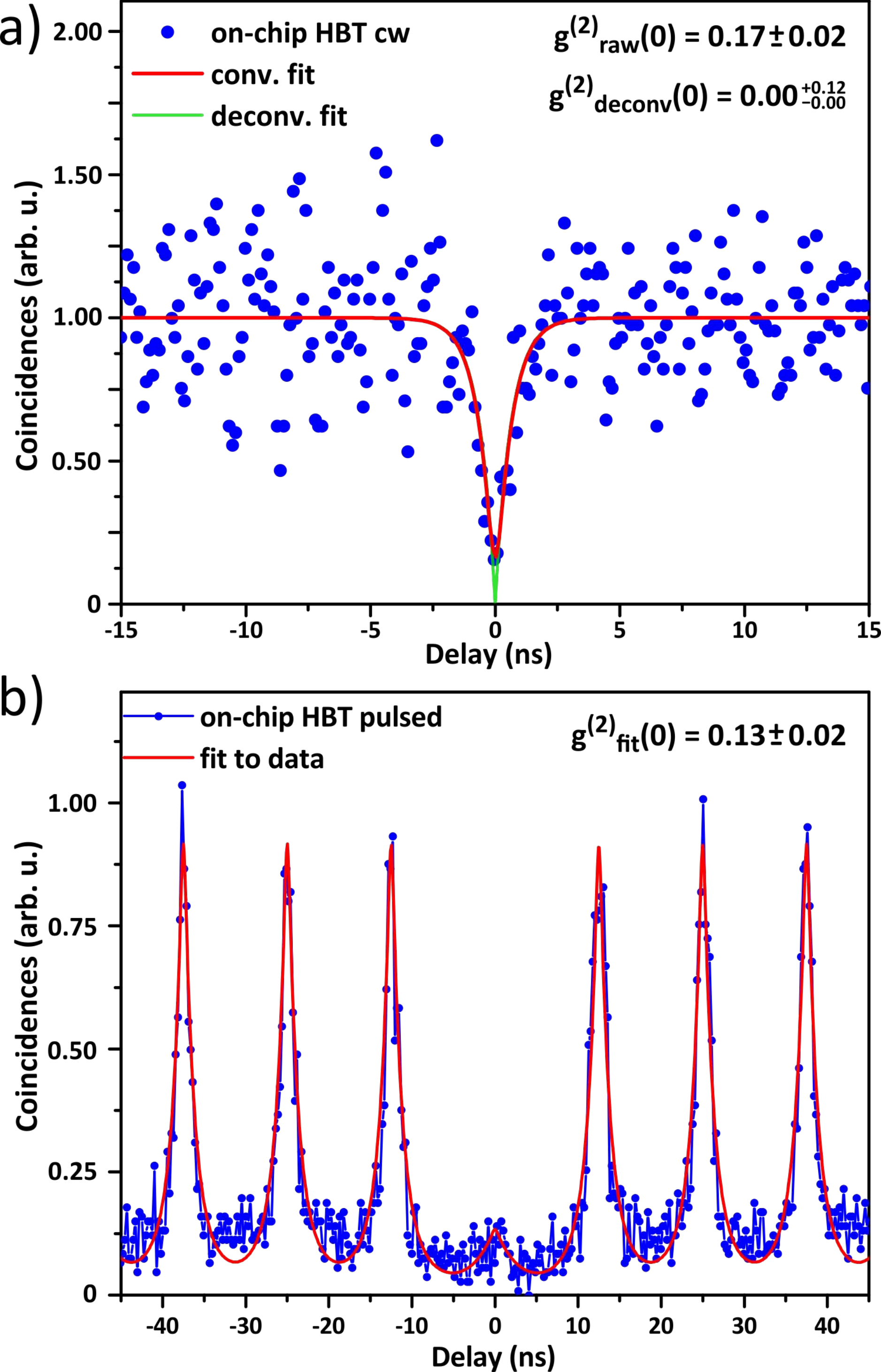}
  \caption{a) Off-resonant, CW excited on-chip HBT measurement of the device shown in Fig.\,\ref{fgr:MMI} a). Deconvoluting the data with the SPCM time resolution yields $g^{(2)}(0)= 0.00^{+0.12}_{-0.00}$. b) On-chip HBT measurement under off-resonant pulsed excitation. Triggered single-photon emission is shown through this on-chip level experiment with $g^{(2)}(0) = 0.13\pm0.02$.}
  \label{fgr:g2}
\end{figure}

A central goal of on-chip quantum photonics is to integrate lab-sized experiments into microchips\cite{Politi2009,Spring2012}. To show the on-chip functionality of our deterministic QD-waveguide-MMI coupler interface, we perform an on-chip HBT experiment (detection off-chip) proving its functionality on a single-photon level\cite{Rengstl2015,Schwartz2016,Prtljaga2014}. For this, we use the device shown in Fig.\,\ref{fgr:MMI}\,a and align the MO to be centered between port 1 and 2, collecting light from both exit ports simultaneously. As the two beams pass through the MO under different angles, they can be coupled into different paths on the optical table using a D-shaped pick-off mirror. Using this technique, a cross-correlation measurement between port 1 and port 2 is performed. Both light beams are spectrally filtered for the $897.3\,$nm line in spectrometers and then directed onto SPCMs. Since the 50/50 beamsplitter is already integrated on the photonic chip, this corresponds to an off-chip HBT autocorrelation measurement. The resulting $g^{(2)}(\tau)$ curves are depicted in Fig.\,\ref{fgr:g2}\,a for CW excitation and in Fig.\,\ref{fgr:g2}\,b for pulsed excitation, giving access to triggered single-photon emission. Deconvoluting the $g^{(2)}(\tau)$ curve for the CW case, we find $g^{(2)}(0)= 0.00^{+0.12}_{-0.00}$. For the pulsed case, we fit $g^{(2)}(\tau)$ with a model function (see \textit{supporting info S4}) and determine $g^{(2)}(0)=0.13\pm0.02$. Thus, triggered single-photon emission is unambiguously observed, proving the functionality of our deterministically integrated on-chip 50/50 splitter. We attribute the non-ideal multi-photon suppression to charge-carrier recapture processes\cite{Aichele2004,Fischbach2017a} and emission from uncorrelated background emitters in the waveguide sections.

In conclusion, we have experimentally demonstrated a scalable single-step fabrication technique for the deterministic integration of spatially and spectrally pre-selected QDs into on-chip quantum photonic device building blocks. Based on this method, we introduced a U-shaped waveguide design for the $\beta$-factor independent characterization of waveguide losses and the MMI coupler transmission. Moreover, the feasibility of MMI couplers for quantum photonic integration was analyzed through simulations and experiments, finding maximum transmissions of $T_{\textrm{Sim}}=97\,\%$ and $T_{\textrm{Exp}}=61\,\%$ for the given material system. 50/50 splitting is confirmed through $\mu$PL spectroscopy in an orthogonal measurement configuration exciting the QD off-resonantly and detecting light from the sample facet. The functionality of the deterministic QD-waveguide-MMI interface is proven by measuring triggered single-photon emission in an on-chip experiment yielding $g^{(2)}(0) = 0.13\pm0.02$. The spatially and spectrally controlled on-chip photonic integration of QDs via in-situ EBL offers a variety of possible applications. For instance, it allows for defining resonators around the QD in a controlled fashion, enabling scalable high-$\beta$ coupling of QD emission into the fundamental waveguide mode\cite{Arcari2014}. Taking into account recent works on hybrid material systems relying on mode conversion for high-$\beta$ coupling\cite{Davanco2017,Wu2017}, deterministic in-situ EBL may ensure a close-to-ideal mode conversion by integrating the QD at the optimum position inside the mode converter. Furthermore, in combination with wavelength fine-tuning concepts\cite{Hallett2017}, the spectrally selective integration of multiple QDs in a single $2\times 2$ beamsplitter paves the way for conducting an on-chip Hong-Ou-Mandel experiment. In another direction, placing two resonant QDs in a single waveguide promises excitonic state transfer between the two quantum emitters.

METHODS

\textit{Sample growth}\\
The QD sample is grown by molecular beam epitaxy on a (100) GaAs substrate in a modified Stranski-Krastanov mode. A GaAs buffer layer is followed by an 1\,$\mu$m Al$_{0.4}$Ga$_{0.6}$As cladding layer, on top of which a 380\,nm GaAs core layer is grown. Low density ($< 10^{7}\, \frac{1}{\textrm{cm}^2}$) InAs QDs are embedded 245\,nm below the sample surface. The sample layout is chosen such that single-mode waveguides can be produced by etching down the GaAs by about 360nm, with the QD layer lying at the maximum of the fundamental TE mode for optimum emission coupling. A 20\,nm thin layer of GaAs remains on top of the AlGaAs cladding to prevent surface oxidation.

\textit{Device fabrication}\\
The sample is coated with 100\,nm CSAR 62 (AR-P 6200) dual-tone EBL resist, which exhibits high contrast, resolution and a favorable onset dose of 12\,$\frac{\textrm{mC}}{\textrm{cm}^2}$ at a temperature of 7\,K\cite{Kaganskiy2016}. Then, it is mounted onto a custom-made liquid helium flow crystat inside a Jeol JSM 840 SEM and cooled down to a temperature of 7\,K. The SEM is operated at a magnification of 300x, a beam current of 0.5\,nA and an acceleration voltage of 20\,kV. These parameters are the best compromise for performing both CL spectroscopy and EBL in immediate succession. The sample is excited by the electron beam and light is collected through an NA$=0.8$ elliptical mirror inside the specimen chamber. Before mapping the sample, the electron beam is positioned 10\,$\mu$m above the designated mapping area to adjust the light-collection optics while not exposing the resist in the area of interest. While scanning over 200\,$\mu$m$^2$ to 400\,$\mu$m$^2$ large regions with a step size of 500\,nm, we record CL spectra using a spectrometer attached to the SEM. The electron dose during mapping is 8\,$\frac{\textrm{mC}}{\textrm{cm}^2}$, which is well below the negative-tone onset-dose. From the maps we extract the positions and spectra of suitable QDs. Subsequently we perform EBL with a self-built pattern generator defining the waveguide and MMI structures by overexposing the CSAR resist with a dose of 50\,$\frac{\textrm{mC}}{\textrm{cm}^2}$. After the patterning the sample is heated up to room temperature. Development of the resist is done by a 45\,s dip in AR-600-546, followed by a stop in isopropyl alcohol and a rinse in de-ionized water. The resist pattern is transferred into the GaAs with an inductively coupled plasma reactive ion etch. The etching is stopped such that a (10\,-\,20)\,nm thin layer of GaAs is left on top of the AlGaAs to prevent oxidization. Finally, residual resist is removed by oxygen plasma ashing and a 24\,h dip in N-methyl-2-pyrrolidone.

\textit{Numerical Simulations}\\
For computing the coupling efficiency $\beta$ of light emitted by the QDs to single-mode waveguide modes we perform simulations using finite-element methods (FEM solver JCMsuite). The QD is modeled as dipole emitter, and Maxwell's equations are solved in a time-harmonic scattering, subtraction-field formulation\cite{Zschiedrich2013} on a 3D computational domain. The coupling efficiency is computed using an overlap integral of the total, scattered electromagnetic field and the field of the guided mode, which is obtained in a separate eigenmode computation, also based on FEM. For computing transmission through the MMI we decompose the geometrical setup into parts, which are invariant in propagation direction, compute the propagating modes in the respective sections and deduce the total 3D field using bidirectional mode propagation. Transmission $T_{\textrm{Sim}}$ is then given by the power flux of the fundamental modes in the two outgoing waveguides normalized to the power flux of the incoming waveguide mode. In the optimization for maximum transmission, the incoming waveguide consists of a material stack as shown in Fig.\,\ref{fgr:sample}\nopagebreak\,a with etch depth $h=360$\,nm and width $w=500\,$nm of the rectangular waveguide section. The width of the incoming and exiting waveguides is linearly tapered to 1040\,nm over a taper length of 10\,$\mu$m. For determining the optimized layout length $L_{\textrm{Sim}}$ and width $W_{\textrm{Sim}}$ of the rectangular MMI section is scanned.

\textit{Optical measurements}\\
Samples are mounted onto a liquid helium flow cryostat with two optical windows, one facing to the top and one to the side of the cryostat. As excitation laser we use a compact 785\,nm CW laser diode for basic characterizations and a Ti:Sapphire laser in either picosecond pulsed or CW mode for the cross-correlation measurements. A 20x MO focuses the laser to a spot of about 3\,$\mu$m diameter on the sample. For detection, a NA$=0.4$ 50x MO is used. The spatial filtering of the 50x MO is strong enough that no pinhole is needed to restrict the collection spot to a single waveguide facet. The light is detected with either a liquid nitrogen cooled CCD camera or fiber-coupled SPCMs with a 250\,ps timing resolution. 

\begin{acknowledgement}
The research leading to these results has received funding from the German Research Foundation through CRC 787 'Semiconductor Nanophotonics: Materials, Models, Devices' and from the European Research Council under the European Unions Seventh Framework ERC Grant Agreement No 615613. KIST authors acknowledge support from KIST's flagship program.
\end{acknowledgement}

\begin{suppinfo}
Details on the MMI geometry, microscope images of the U-shaped waveguide-MMI structures for MMI transmission measurements, analysis of the MMI transmission dependent on the length $L$, model for pulsed $g^{(2)}(\tau)$ measurements
\end{suppinfo}



\providecommand{\latin}[1]{#1}
\makeatletter
\providecommand{\doi}
  {\begingroup\let\do\@makeother\dospecials
  \catcode`\{=1 \catcode`\}=2 \doi@aux}
\providecommand{\doi@aux}[1]{\endgroup\texttt{#1}}
\makeatother
\providecommand*\mcitethebibliography{\thebibliography}
\csname @ifundefined\endcsname{endmcitethebibliography}
  {\let\endmcitethebibliography\endthebibliography}{}

\newpage

\end{document}